\documentclass{aa}  

\usepackage{graphicx}
\usepackage{natbib}
\usepackage{txfonts}

\begin{document}

   \title{On the origin of two unidentified radio/X-ray sources discovered with XMM-{\em Newton}}
   
   \author{Federico Garc\'{\i}a \inst{1,2,}\thanks{Fellow of CONICET, Argentina.}, Jorge A. Combi\inst{1,2}, Mar\'{\i}a C. Medina\inst{1}, \and Gustavo E. Romero\inst{1,2}}

   \authorrunning{F. Garc\'{\i}a et al.}

\titlerunning{On the origin of two unidentified radio/X-ray sources discovered with XMM-{\em Newton}} 

\offprints{F. Garc\'{\i}a}  

   \institute{Instituto Argentino de Radioastronom\'{\i}a (CCT La Plata, CONICET), C.C.5, (1894) Villa Elisa, Buenos Aires, Argentina\\
   \email{[fgarcia,jcombi,clementina,romero]@iar-conicet.gov.ar}
   \and
   Facultad de Ciencias Astron\'omicas y Geof\'{\i}sicas, Universidad Nacional de La Plata, Paseo del Bosque, B1900FWA La Plata, Argentina}

   \date{Received; accepted}

 
\abstract
{}
{We aim at clarifying the nature of the emission of two spatially related unidentified X-ray sources detected with XMM$-${\it Newton} telescope at intermediate-low Galactic latitude}
{We use the imaging and spectral capabilities of XMM$-${\it Newton} to study the X-ray properties of these two sources. In addition, we complement our study with radio data obtained at different frequencies to analyze a possible physical association between the sources.}
{Observations reveal a point-like source aligned with elongated diffuse emission. The X-ray spectra of these sources is best-fitted by an absorbed power law with photon index $\Gamma\sim1.7$ for the point-like source and $\sim$2.0 for the extended source. Both sources show nonthermal radio-continuum counterparts that might indicate a physical association. In addition, from the available data, we did not detect variability on the point-like source in several timescales. Two possible scenarios are analyzed: one Galatic and one extra-Galatic. First, based on HI line absorption, assuming a Galactic origin, we infer a distance upper bound of $\lesssim$2~kpc, which poses a constraint on the height over the Galactic plane of $\lesssim$200~pc and on the linear size of the system of $\lesssim$2.3~pc. In this case, the X-ray luminosities are $\gtrsim$10$^{32}$~erg~s$^{-1}$ and $\gtrsim$7.5$\times$10$^{32}$~erg~s$^{-1}$, for the point-like and extended sources, respectively. Second, an extra-Galactic nature is discussed, where the point-like source might be the core of a radio galaxy and the extended source its lobe. In this case, we compare derived fluxes, spectral indices, and spatial correlation with those typical from the radio galaxy population, showing the feasibility of this alternative astrophysical scenario.}
{From the available observational evidence, we suggest that the most promising scenario to explain the nature of these sources is a system consisting of a one-sided radio galaxy, where the point-like source is an active galactic nucleus and the extended source corresponds to the emission from its lobe. Other possibilities include a pulsar and its associated pulsar wind nebula, where the radio/X-ray emission originates from the synchrotron cooling of relativistic particles in the pulsar's magnetic field or a casual alignment between two unrelated sources, such as an active galactic nucleus and a Galactic X-ray blob. Deeper dedicated observations in both radio and X-ray bands are needed to fully understand the nature of these sources.}
   

   \maketitle

\section{Introduction}

\begin{figure*}
\centering
\includegraphics[width=9.0cm,angle=0]{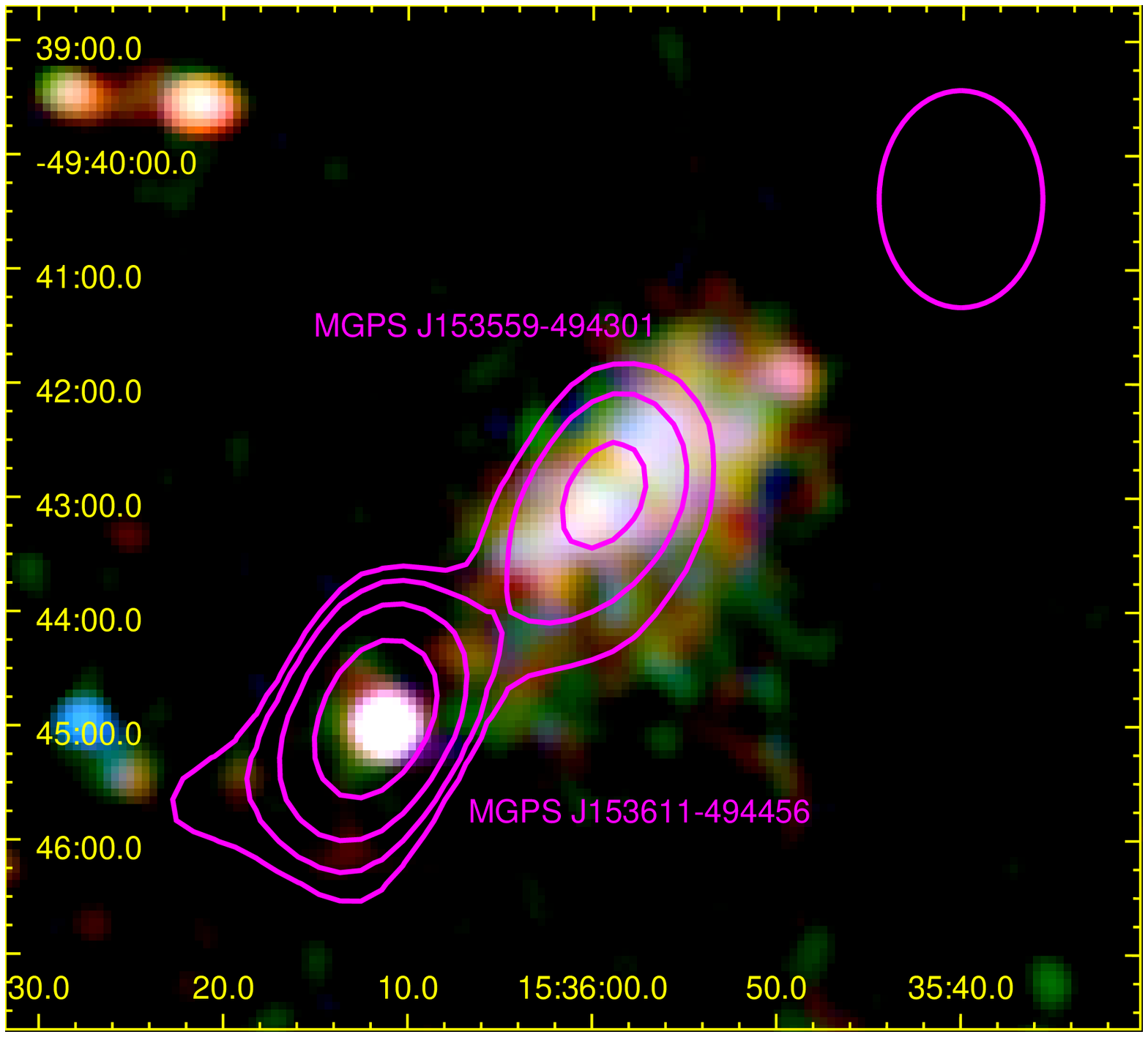}
\includegraphics[width=9.0cm,angle=0]{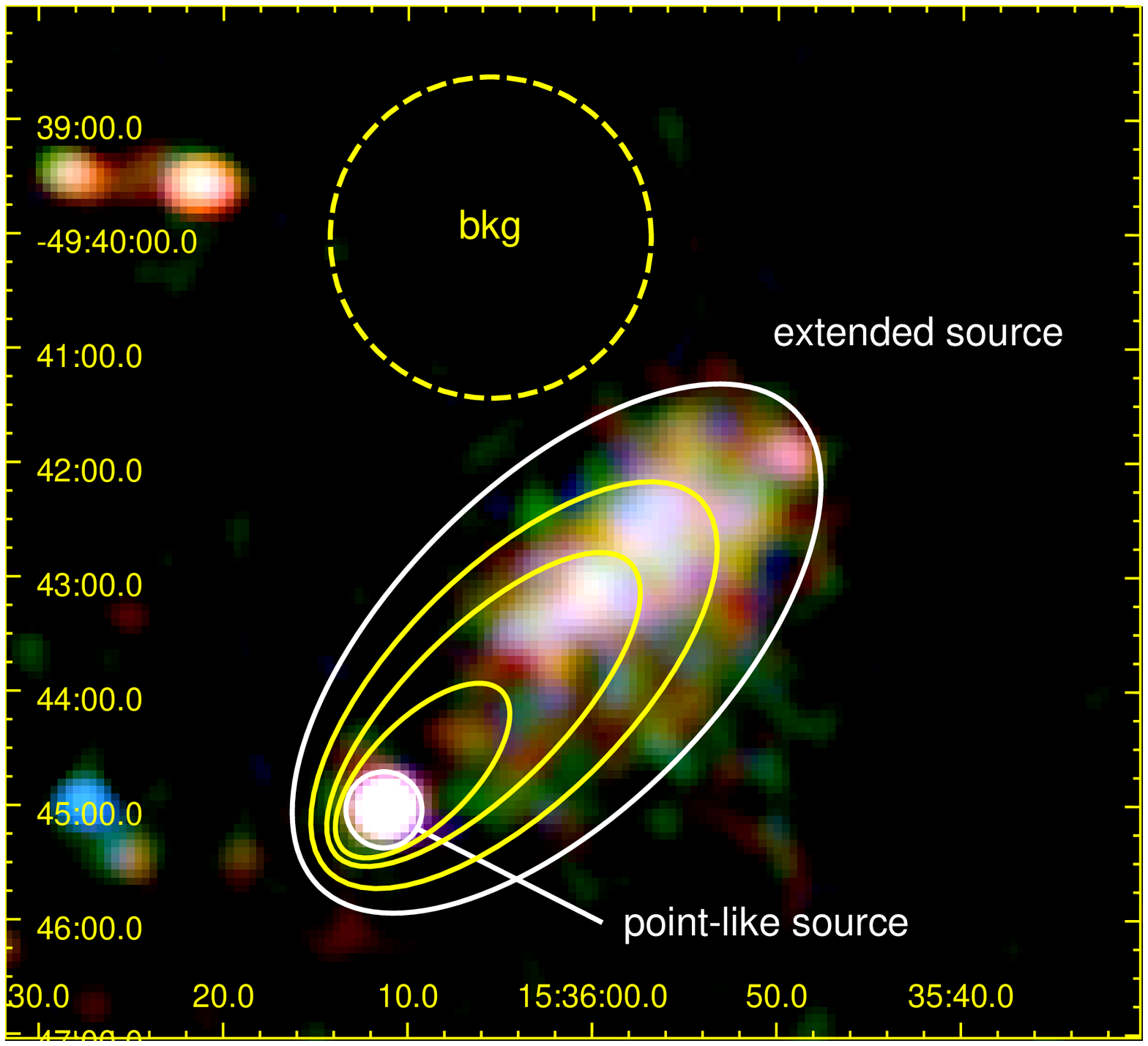}
\caption{XMM--{\it Newton} X-ray image of the emission in the 0.5--8.0 keV energy range of the extended source. Soft X-rays photons (0.5--2.0 keV) are shown in red, medium X-ray photons (2.0--3.5 keV) in green, and hard X-ray photons (3.5--8 keV) in blue. {\bf Left panel:}  radio contours at 843 MHz are overlaid in steps of 0.01, 0.03, 0.07, and 0.18 mJy beam$^{-1}$ in magenta, and the nominal beam ellipse of 43 $\times$ 57 arcsec are indicated in the
upper right corner in magenta. {\bf Right panel:} X-ray spectra extracting regions are indicated with white and yellow ellipses. Background region is indicated with a yellow dashed line.}
\label{images}
\end{figure*}

With the advent of the new generation of high-sensitivity X-ray observatories, such as XMM-Newton, which operates in the 0.2–12 keV energy range, a large number of point-like and extended unidentified X-ray sources were discovered at middle and high Galactic latitudes \citep{watson2009}. An instrument with this kind of capability offers a good opportunity to detect and investigate X-ray emission from distant and obscured high-energy sources. These sources range from undetected pulsars, accreting black holes or supernova remnants (SNRs), to active galactic nuclei (AGNs).
A system composed of a point-like source and extended X-ray emission with comet-like structure and nonthermal radio/X-ray emission could have two possible origins, either Galactic or extra-Galactic.

In our Galaxy, this kind of system can be the result of a pulsar moving at supersonic velocity with respect to its surroundings \citep{bucciantini2001,vanderswaluw2003}. In this case, the radio/X-ray emission originates from synchrotron radiation of the high-energy particles injected by a spinning neutron star through its own magnetic field, namely, a pulsar wind nebula (PWN). The $\sim$50 X-ray PWNe detected so far \citep{kp2008} show several different observational properties: X-ray spectral indices in the 1--2 range, X-ray luminosities ranging from 10$^{31}$ to 10$^{36}$~erg~s$^{-1}$, and radio/X-ray morphologies going from regular, clearly toroidal or bow-shock-like shapes in young SNRs to irregular shapes in older systems \citep[for a detailed review, see][]{gaenslerslane2006}. In some cases, because of a density gradient in the interstellar medium (ISM), the PWN can detach from the pulsar, resulting in a relic PWN \citep[e.g.,][]{vanderswaluw2004}. Althought most of PWNe had been associated with pulsars and/or SNRs, some of them remain candidates because of the lack of conclusive observational evidence \citep{matheson2013,kargaltsev2013}. 

In an extra-Galactic scenario, these systems can be associated with AGNs. Possible candidates are FR I radio galaxies \citep{beckmann-shrader2012}, where nonthermal radio and X-ray emission is produced by charged particles accelerated within the jet. The broadband spectra (from UV to X-rays) of their jets can be fitted by a simple synchrotron model \citep[e.g., 3C 66B,][]{hardcastle2001} with a steep X-ray component. However, similar X-ray morphology has also been found in several BL~Lac objects; e.g., AP~Lib \citep{kwt2013}, S5 2007+777 \citep{sambruna2008}, OJ~287 \citep{marscherjorstad2011}, PKS 0521--365 \citep{birkinshaw2002}, 3C371, and PKS 2201+044 \citep{sambruna2007}. The spectra of the jets in these objects are generally flat and they are considered to be inverse-Compton (IC) dominated \citep[synchrotron self Compton or the result of scattering of cosmic microwave background photons,][]{harris2002}. There is a correlation between the jet and the core spectral indices, which means that IC radiation should also dominate the core X-ray emission. The core of these BL Lac objects has been detected by Fermi, presenting softer spectra in gamma-rays (spectral index ranging from 2 to 2.4), opening the possibility that the IC component could reach  TeV energies.

In this paper, we report a radio/X-ray study of two possibly associated unidentified X-ray sources. One source is point-like and the other has a diffuse and elongated morphology, detected with the XMM-Newton Observatory. We analyze the two possibilities on the origin of
this emission (Galactic or extra-Galactic) mentioned above. The structure of our paper is as follows: in Sect. 2 we describe the observations and the data reduction process. X-ray analysis of the sources is presented in Sect. 3. In Sect. 4, we discuss the results found in the X-ray band together with those coming from the radio analysis of pre-existent data, and we suggest some possible astrophysical explanations for the sources, either in the context of Galactic or an extra-Galactic origin. Finally, in Sect. 5, we summarize our main conclusions.

\section{X-ray observations and data reduction}

The point-like and extended X-ray sources under study in this work were observed on August 2012 by the XMM-{\it Newton} X-ray observatory (ObsID. 0692830301). The observation was pointed toward the unidentified $\gamma$-ray source {\mbox 2FGL J1536.4--4949} ($\alpha_{\rm J2000.0}$=$15^{\rm h} 36^{\rm m} 29\fs8 $, $\delta_{\rm J2000.0}$=$-49\degr 49\arcmin 45\farcs9 $) with a 68\% error ellipse of $1.69 \times 1.58$~arcmin and an inclination of $90\degr20\arcmin$ with respect to north. Data was acquired with the European Photon Imaging Camera (EPIC) by means of MOS \citep{turner2001} and PN \citep{struder2001} detectors, operating in the 0.2--15~keV energy range, and for a net exposure time of 17.91~ks. The observation was performed using a {\it \textup{medium}} filter for MOS1/2 and a {\it \textup{thin}} filter for PN in the full frame (FF) imaging mode, with timing resolutions of 2.5~s and 200~ms for MOS and PN detectors, respectively\footnote{http://XMM-Newton.vilspa.esa.es/externalXMM-Newton\_user\_support/documentation/}. 

We obtained raw data of this observation from the XMM-{\it Newton} Science Archive (XSA)\footnote{http://XMM-Newton.vilspa.esa.es/xsa/}, which we calibrated using XMM-{\it Newton} Science Analysis System (SAS)\footnote{http://XMM-Newton.vilspa.esa.es/externalXMM-Newton\_sw\_cal/sas.shtml} version 13.5.0 and latest calibration files. We extracted light curves of photons above 10~keV for the entire field of view of the EPIC cameras and discarded high-background time intervals to produce a Good Time Interval (GTI) file in order to exclude periods of high background
due to soft-proton flares, which could eventually affect the observations. We ended up with a total live time of 17.44~ks, 17.45~ks, and 17.82~ks for the MOS-1, MOS-2, and PN cameras, respectively. To create  spectra and light curves images, we selected events with {\sc flag 0}, and {\sc patterns}$\leq$ 12 and 4 for MOS and PN cameras, respectively. 

After a quick inspection of the datasets we found two unidentified, spatially associated X-ray sources in the northwest part of the detector: a point-like source and an extended source. Both sources had been recently included in the 3XMM-DR5 catalog \citep{3xmmdr5}. According to this catalog, the point-like source 3XMM J153611.2--494500 is located at ($\alpha_{\rm J2000.0}$=$15^{\rm h} 36^{\rm m} 11\fs 2$, $\delta_{\rm J2000.0}$=$-49\degr 45\arcmin 00\arcsec$) with a position error of $0.58\arcsec$ and the extended source 3XMM J153559.6-494306 is centered at ($\alpha_{\rm J2000.0}$=$15^{\rm h}35^{\rm m}59\fs6$, $\delta_{\rm J2000.0}$=$-49\degr43\arcmin06\arcsec$) elongated $\sim$4$\arcmin$~in the northwest-southeast direction with a position error of $1.86\arcsec$. The off-angle of both sources with respect to the pointing of the observation is about $\sim$5.5\arcmin and 8\arcmin, respectively, which implies a reduction of the effective area of about 9\% and 11\%, respectively. In which follows, we focus our work on the study of these two previously unindentified X-ray sources.\\

\section{Results}

\subsection {X-ray images}

We produced X-ray images for MOS and PN cameras that we combined  to increase the signal-to-noise ratio (S/N) by means of the {\sc emosaic} SAS task. We prepared the corresponding set of exposure maps for each camera to account for spatial quantum efficiency and mirror vignetting by running the SAS task {\sc eexmap}. We performed exposure vignetting corrections dividing each image by its corresponding superimposed exposure map. Finally, we smoothed the images with a three-pixel Gaussian filter.

In Fig.~\ref{images} we show the resulting narrowband X-ray images of the region for soft (0.5--2.0 keV), medium (2.0--3.5 keV), and hard (3.5--8 keV) energy bands in red, green, and blue, respectively. Above 8.0~keV, no significant X-ray emission was detected. In the left panel, radio contours at 843 MHz \citep{whitegreen} from the Molonglo Observatory Synthesis Telescope (MOST) were superimposed in magenta, and  a nominal beam size of 43~$\times$~57 arcsec is indicated in the upper right corner. In the right panel, we overlaid X-ray spectra extraction regions. We used solid lines to indicate source regions and dashed line for the background. In the images, north is up and east is to the left. As it can be easily observed, a point-like source (hereafter Source~1) seems to be connected with an elongated diffuse X-ray source (hereafter Source~2), which extends along the southeast/northwest axis. Both X-ray sources spatially correlate very well with a continuum radio source with two peaks: Source~1 with {\mbox MGPS J153611--494456} and Source~2 with {\mbox MGPS J153559--494301} \citep{murphy2007}. The extended X-ray emission is elongated along a position angle of approx. -45$^{\circ}$, with an apparent angular size of roughly 1.5 $\times$ 4~arcmin on the plane of the sky, being energy independent as the shape of the X-ray Source 2 does not noticeably change in the color-coded image.

\subsection{X-ray spectral analysis}

\begin{figure}
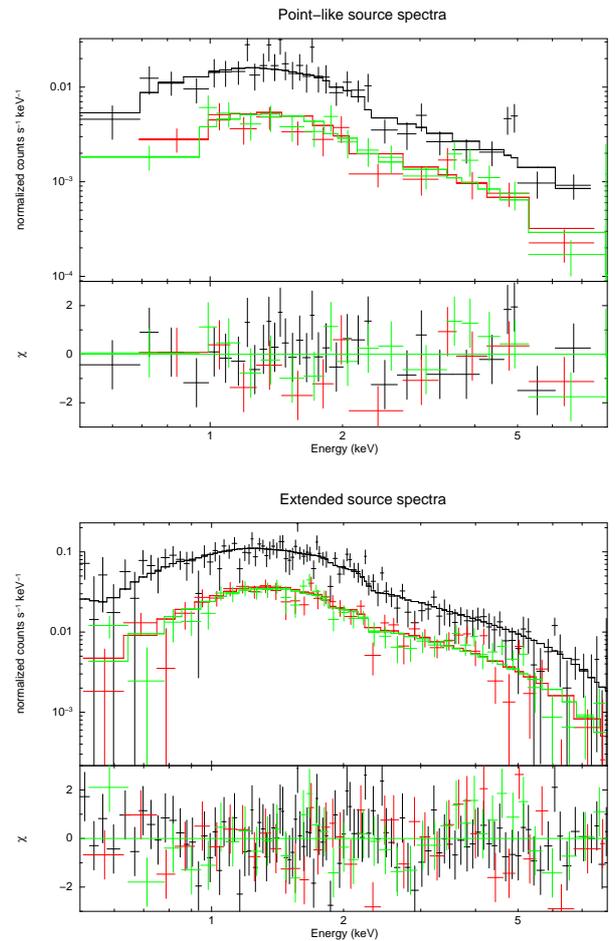


\includegraphics[width=6cm,angle=270]{Nellipse0.ps}\\

\includegraphics[width=6cm,angle=270]{Nellipse5.ps}

\caption{XMM-{\it Newton} PN (black) MOS1 (red), and MOS2 (green) spectra of the point-like Source~1 (upper panel) and the extended Source~2 (lower panel). Solid lines indicate the best-fit, absorbed power-law model (see Table~1). Lower panels show $\chi^{2}$ fit residuals.}
\label{spectra}
\end{figure}

\begin{table}
\caption{Spectral parameters of the X-ray emission of point-like Source~1 and extended Source~2.}
\label{table1}
\renewcommand{\tabcolsep}{0.3cm}
\begin{centering}
\begin{tabular}{l| cc}
\hline\hline
Model \& Regions & Source~1 & Source~2 \\
\hline
TBABS*POWERLAW & & \\
N$_\mathrm{H}$ [10$^{22}$~cm$^{-2}$] &  0.38$\pm$0.06 & 0.46$\pm$0.04  \\
$\Gamma$    &           1.73$\pm$0.11   & 1.96$\pm$0.08   \\
\hline
$\chi^{2}_{\nu}$ / d.o.f. & 1.02/145  & 0.97/877  \\
\hline
Flux(0.5$-$8.0~keV) & 0.21$\pm$0.01 &  1.71$\pm$0.07
\\
\hline
\end{tabular}
\tablefoot{Error values are 1-$\sigma$ confidence levels for every single parameter. Fluxes are absorption corrected and they are given in units of 10$^{-12}$~erg~cm$^{-2}$~s$^{-1}$.}
\end{centering}
\end{table}

In order to analyze the physical conditions of the X-ray emission of these sources, we first extracted spectra from a circular region (PS) for Source~1 and an elliptical region (excluding PS) for Source~2. We indicated both regions in the right panel of Fig.~\ref{images} with white solid lines. Ancillary response files (ARFs) and redistribution matrix files (RMFs) were produced by means of the {\sc arfgen} and {\sc rmfgen} SAS tasks, respectively. In the case of PS, the point spread function (PSF) correction was applied. Background was subtracted using the dashed line region shown in the same figure. We performed the spectral analysis via the XSPEC package \citep{arnaud1996} working in the 0.5--8.0~keV energy range because no significant X-ray emission was detected above this limit.

\begin{table*}
\begin{centering}
\caption{Spectral parameters of the X-ray emission of the selected elliptical regions.}
\label{table2}
\renewcommand{\tabcolsep}{0.2cm}
\begin{tabular}{l | c c c c c}
\hline\hline
Model \& Regions  & PS & $E_1$ & $E_2$ & $E_3$ & $E_4$\\
\hline
TBABS*POWERLAW\\
N$_{\rm H}$ {[}10$^{22}$~cm$^{-2}${]}  & 0.38$\pm$0.06 & 0.26$\pm$0.14 & 0.48$\pm$0.06 & 0.51$\pm$0.06  & 0.47$\pm$0.07 \\
$\Gamma$   & 1.73$\pm$0.11  & 1.85$\pm$0.33  & 1.92$\pm$0.11 & 1.99$\pm$0.11  & 2.00$\pm$0.13 \\
\hline
$\chi^{2}_{\nu}$ / d.o.f. & 1.02/145  & 0.69/71 & 1.07/281 & 0.99/325 & 0.98/404 \\
\hline
Flux (0.5--8.0 keV) & 0.21$\pm$0.01 & 0.06$\pm$0.01 & 0.49$\pm$0.03 & 0.56$\pm$0.06 & 0.63$\pm$0.04 \\
\hline                                                                   
\end{tabular}
\tablefoot{Error values are 1-$\sigma$ confidence levels for each parameter. Fluxes are absorptioncorrected and they are given in units of 10$^{-12}$~erg~cm$^{-2}$~s$^{-1}$.}
\end{centering}
\end{table*}

We fitted EPIC spectra with a single nonthermal model, described by a power law, affected by an interstellar-absorption component \citep[{\sc tbabs};][]{tbabs}. The goodness of the model fit was derived according to the $\chi^2$-test statistics. In Fig.~\ref{spectra} we show the background subtracted X-ray spectra of Source~1 and 2 in upper and lower, panels, respectively. In solid lines we plot the best-fit model obtained by fitting simultaneously PN (black) and MOS1/2 (red and green, respectively) data. The lower panels show $\chi^2$ fit residuals. The parameters of the best fit to the spectra of Source~1 and 2 are presented in Table~1, where errors are quoted at 1-$\sigma$ confidence limits. We tried to fit the spectra using a single thermal blackbody component, but we were not able to obtain a good fit (reduced $\chi_\nu^2=1.3$ for $N_{\rm H} \sim 0$).  We added a thermal component to the nonthermal best-fit model to constrain the thermal emission coming from the point source. This did not improve the fit significantly, since the fits obtained for a single nonthermal component already yielded  $\chi^2 \approx 1$. Nevertheless, we obtained an upper limit $\lesssim$15\% for the unabsorbed thermal flux detected in the 0.5--8.0~keV energy range, using a black body with its temperature fixed at the best-fit value $kT = 0.14$~keV.

To better understand the physical structure of the diffuse emission, we also performed spatially resolved X-ray spectroscopy. For this purpose, following \citet{bocchino2001} and \citet{gaensler2004}, we split the elongated emission into several increasing elliptical annuli, $E_{\rm i}$ for $i$=$1 ... 4$ , running from the circular PS region (coincident with Source~1) to region $E_4$ whose outer ellipse coincides with the white ellipse of the region Source~2(shown in yellow in the right panel of Fig.~\ref{images}). Spectra was produced and fitted with the same approach and data was again best fitted with a single power-law component. In Table~2 we show the best-fit parameters obtained in our analysis.

According to the results presented in Tables 1 and 2, Source~1 appears harder ($\Gamma \sim 1.7$) than Source~2 ($\Gamma \sim 1.96$), whereas the absorption columns of both regions are consistent within the errors. This result is confirmed by the detailed, spatially resolved spectroscopy analysis. To avoid the effects caused by variations found in the absorption column for the best-fit parameters, we also performed the same analysis but keeping $N_{\rm H}$ fixed to the best-fit value of Source~1. Following this approach, we found consistent results both in the spectral indices and  total flux. The spectrum of $E_1$ region could possibly be contaminated by the residual PSF of the point-like source, and while regions $E_2$ to $E_4$ show an increasing power-law index with the distance to the point-like source, all of the spectral indices are consistent within the error bars.

To obtain a statistical assessment of possible X-ray pulsations of Source~1, we used the 17.8~ks EPIC-PN observation to compare the time arrival distribution of source photons. We used the PS extraction region centered in the point-like source, which gave us a total of $\sim$2000 counts, giving an average count rate of roughly 0.5~counts/s. We produced the FFT of the light curve via {\sc powspec} with different approaches for both unbinned and binned data and we found no significant peak in the resulting power spectra. We also performed a pulsation search based on an H-test \citep{dejager1989} on the phase of unbinned events. For an independent period search of $10^5$ trials over the Nyquist limit, we found one peak with an H-probability $\lesssim$10$^{-5}$, which is fully compatible with a by-chance detection for the number of trials used. Thus, we discarded pulsations with a period greater than twice the read-out time of the EPIC-PN camera in the FF mode, which corresponds to $\sim$0.4~s.

We also searched for variability on the XMM-Newton observation at 10s and 100s timescales. We performed a $\chi$-square test via {\sc lcstats} and we found no significant variability. In addition, we analyzed three observations available in the {\it Swift}\footnote{http://swift.gsfc.nasa.gov/archive/} archive from June and August 2009 and January 2010 of 1000, 750 and 5200~s exposures, respectively. From data adquired in the photon-counting mode, using {\sc xselect} task, we obtained the count rate of Source 1 in the 0.5--10~keV energy band, resulting in 0.011$\pm$0.003, 0.007$\pm$0.003 and 0.008$\pm$0.001~counts~s$^{-1}$, respectively, which also discards variability at 1-$\sigma$ confidence levels at these timescales. Finally, we compared {\it Swift} count rates with XMM-{\it Newton} spectral results (that we converted to expected count rates via {\sc PIMMS}), resulting consistent between each other within the errors. Hence, no clear evidence of variability was detected at any timescale.

\subsection{Fermi $\gamma$-ray data analysis }

We used data from the {\it Fermi}\footnote{http://fermi.gsfc.nasa.gov/} Large Area Telescope (LAT) to search for putative $\gamma$-ray emission from these sources in a region centered at ($\alpha$=$233.98^{\circ}$, $\delta$=$-49.71^{\circ}$) within a circle of 3$^{\circ}$ of radius of interest (ROI). The data analysis was performed using the Science Tools package (v9r32p5) available from the Fermi Science Support Center (FSSC). Data were obtained from the reprocessed Fermi Pass~7 database and the instrumental response function {\sc P7REP SOURCE V15} was used. We analyzed six years of data, from August 2008 to August 2014 (2008-08-04T15:43:36 to 2014-08-28T08:34:21, UTC). To prevent  contamination by the Earth's albedo, the events with zenith angle greater than 100$^{\circ}$ or rocking angle greater than 52$^{\circ}$ were filtered. We performed unbinned likelihood analysis via the {\sc gtlike} task. To model background source contributions, we included all two-year {\it Fermi} Gamma-ray LAT (2FGL) point sources \citep{Nolan2012} associated with the extended source templates within 3$^{\circ}$ from the ROI center, including the unidentified $\gamma$-ray source {\mbox 2FGL J1536.4--4949}, possibly associated with the recently discovered millisecond pulsar MSP J1536--4948 \citep{ray2012}, together with the Galactic diffuse background (gll\_iem\_05) and the isotropic diffuse background (iso\_source\_v05). We modeled the differential $\gamma$-ray flux expected from a point source located at the center of our ROI with a simple power law. We did not detect significant emission in the chosen energy band at the location of the X-rays sources, so only an upper limit of 3.53$\times$10$^{-10}$~ph~cm$^{-2}$~s$^{-1}$ for the 3--30~GeV energy range could be imposed.

\section {Discussion}

The morphological and spectral information gathered from radio and X-ray frequencies about the compact point-like Source~1 and the elongated extended Source~2 can be interpreted in at least three different astrophysical scenarios. Assuming a physical association between them, in a Galactic setting we could be observing a system composed of a pulsar and its associated PWN, which is moving at supersonic velocity with respect to its surroundings. In an extra-Galactic context, the system could be a radio galaxy composed of an AGN and an extended one-sided lobe. Finally, if the physical connection is not real, we might just be observing a casual alignment between a background AGN and a Galactic X-ray blob.

Concerning our X-ray spectral analysis, we found that both the point-like and  diffuse sources are well fitted by a single power law with photon indices $\Gamma$=1.7$\pm$0.1 and $\Gamma$=1.96$\pm$0.08, respectively, while thermal models do not fit the data well. Best-fit absorption columns of both sources are consistent within the errors, and also consistent with the value derived in the radio wavelengths \citep[$N_{\rm H}=3\times10^{21}$~cm$^{-2}$,][]{mcclure2009,kalberla2010}. Moreover, no emission lines are present in the spectra, which would be a clear indication of a thermal plasma origin. We also looked at data from 2MASS\footnote{http://www.ipac.caltech.edu/2mass/}, but we did not find infrared counterparts for any of the sources. Thus, the observational evidence points to a nonthermal origin for the X-ray emission of both sources. 

Since the beam-sizes of radio-continuum data available in the literature are very different from each other, we can only roughly estimate a mean spectral index for the whole system (the point-like plus the extended source). From the available radio fluxes of 1230$\pm$40~mJy at 408~MHz \citep{large1981}, with a nominal beam-width of $2.62 \times 4.81$~arcmin, 883$\pm$16.7~mJy at 843~MHz \citep{murphy2007}, with a beam size of $43 \times 57$~arcsec (including both sources), and 156$\pm$12~mJy at 4850~MHz \citep{gregory1994}, with a beam size of 4.8~arcmin, we derive a mean spectral index of $\alpha \sim -0.78$, confirming a nonthermal origin of the radio sources.

The high value of radio flux at 408~MHz and the spectral index are typical of an AGN. If the mean spectral index comes from a combination of the emission of a pulsar/PWN system (roughly -1.8 and -0.3--0.0, respectively), a high fraction of the flux at 408~MHz has to be assigned to the pulsar. In this case, it would be very hard to explain how it was not detected yet in any Parkes pulsar survey. Despite that in our X-ray analysis we could not find pulsations with high confidence, and the high Nyquist limit of 0.4~s does not allow us to discard this possibility.

As we mentioned above, in the extra-Galactic case if there is a physical association between both sources, the system might be a one-sided radio galaxy composed of a core  (the point-like source) and a jet (the extended source). Generally, the core is a compact component, unresolved when  observed with low angular resolution $\geq$ 0.1 arcsec, and coincident with an associated optical object. The core's contribution to the total radio luminosity varies from 1~\% to almost 100~\% in some quasars, a fact that can be interpreted in terms of a beaming model that unifies radio galaxies and quasars into a single population. On the other hand, the lobes are extended regions of radio emission approximately, symmetrically placed on opposite sides of the central core. The overall size of the radio structure can extend for several hundred kpc and in some extreme cases up to a few Mpc. In some quasars, an extended structure only occurs  on one side of the nucleus. One of the reasons for one-sidedness could be that the double structure is oriented close to the line of sight. In our case, from the radio fluxes observed at 843~MHz, where the sources can be resolved, we suggest that the core and the extended source could be classified as a one-sided FR I radio galaxy, which is supported by the mean spectral index of $S \propto \nu^{-0.78}$ estimated at the radio frequencies. An optical/IR counterpart was not detected; this may be because of the limited resolution of the available data and the low Galactic latitude where the source is located. Depending on the jet and ambient medium parameters, most radio galaxies display sizes below 100 kpc \citep{begelman1984}. If this is the case, using standard Friedmann-Lema\^itre-Robertson-Walker formulae, the $\sim$4\arcmin size of our extended source translates into a distance smaller than 80 Mpc, which yields radio luminosities of $1.1\times10^{23}$~W~Hz$^{-1}$~sr$^{-1}$ at 4590~MHz and an X-ray flux at $3.1\times10^{16}$~W~Hz$^{-1}$~sr$^{-1}$ at 2~keV, in agreement with \citet{worrall}.

In the case of a Galactic origin, if both sources are physically connected, it is possible to constrain their distance with \ion{H}{i} data from the Parkes Galactic All-Sky Survey \citep[GASS;][]{mcclure2009,kalberla2010} with a nominal angular resolution of $\sim$16~arcmin to analyze the \ion{H}{i} line intensity of the medium pointing to the direction of these sources and to search for possible absorption features caused by the observed radio-continuum emission. From the velocity profile of the \ion{H}{i} line in the $-400$ to $+400$~km~s$^{-1}$ range, we found three minima at velocities of $-39$, $-12,$ and $-2.6$~km~s$^{-1}$. Based on the Galactic rotation curve obtained by \citet{fich1989}, these velocities correspond to heliocentric distances of $\sim$3 or $\sim$11~kpc, for $-39$~km~s$^{-1}$ and $\la$2 or $\gtrsim$13~kpc for the other cases. An eye inspection of the \ion{H}{i} datacube suggests that the minimum in the profile at $-12$~km~s$^{-1}$ can be an absorption feature, whereas the minima in the other cases is related to bright clouds in the eastern side of the maps. Based on the flux and angular size of the diffuse source, distances $\gtrsim$10~kpc can be safely discarded for a Galactic origin. Hence, we suggest that the radio/X-ray sources should be at a distance $\la$2~kpc, which places a constraint on their height over the Galactic plane in $\la$200~pc and on its linear size of $\gtrsim$2.3~pc, which is compatible with observed PWN sizes \citep{gaenslerslane2006,kp2008}. Moreover, based on this distance, from our X-ray spectral fits we infer X-ray luminosities in the 0.5--8~keV energy band of $\gtrsim$10$^{32}$~erg~s$^{-1}$ and $\gtrsim$7.5$\times$10$^{32}$~erg~s$^{-1}$ for the point-like Source 1 and  extended Source~2, respectively. These sources are  compatible with typical luminosities of pulsar/PWN associations \citep[see lower panel of Fig. 5 from][]{kp2008}. Following the comparison with the pulsar/PWN population, the power-law indices $\Gamma$ of both the putative pulsar and its PWN fall in the typical 1.0--2.0 range. However, the lack of pulsations together with the high radio flux make this scenario rather unlikely.

The unidentified $\gamma$-ray source {\mbox 2FGL J1536.4--4949} located at the center of the XMM-{\it Newton} observation field of view has been detected by {\it Fermi}-LAT with a significance of 40-$\sigma$ and a flux of 1.2$\times$10$^{-8}$~ph~cm$^{-2}$~s$^{-1}$ in the 1 and 100~GeV energy range. Sources 1 and 2 lie about 0.15$^\circ$ outside the 95$\%$ confidence limit of the $\gamma$-ray source, which makes a physical association unlikely. Moreover, in the position where the radio/X-ray sources lie, only an upper limit of 3.53$\times$10$^{-10}$~ph~cm$^{-2}$~s$^{-1}$ for the 3--30~GeV energy range could be derived. 

\section {Conclusions}

We have conducted a morphological and spectral analysis of two spatially associated unidentified X-ray sources (a point-like  and a diffuse-extended source) based on an XMM-{\it Newton} observation and focused in the 0.5--8.0~keV energy range. From the information gathered, the system can be interpreted in at least three different ways. Assuming a physical association, if the system is Galactic, we might be detecting a pulsar/PWN pair. In an extra-Galactic scenario, the system may be a radio galaxy composed of an AGN and an extended one-sided jet. Otherwise, if there is no physical connection, we might just be observing a casual alignment between a background AGN and an elongated Galactic X-ray blob. In this case, the nature of the extended source would remain unsolved.

In the case of an extra-Galactic origin, the most likely configuration is that of a FR I radio galaxy with a single side outflow. The X-rays and radio data for the extended emission of FR I galaxies is generally explained by synchrotron emission from a single population of relativistic electrons gyrating in a  magnetic field inside the jet. This is supported by the fact that the  spectral index of the X-rays is > 1 and significantly larger than the radio index. Also, the morphologies in radio and X-rays of the extended emission are related \citep{harris2006}. The lack of detection in optical or IR bands (likely due to the proximity of the source to the Galactic plane) makes it difficult to go further into this interpretation. However taking into account  the synchrotron model for the observed radiation, no detection at high energies should be expected as a cutoff must appear at harder X-rays. 

For a Galactic origin, based on the \ion{H}{i} data available, we suggest that the radio/X-ray system might be at a distance $\la$2~kpc, posing a constraint on its height over the Galactic plane of $\lesssim$200~pc and on its linear size in $\lesssim$2.3~pc, which are compatible with observed PWN sizes \citep{gaenslerslane2006,kp2008}. Moreover, the inferred X-ray luminosities in the 0.5--8~keV energy band of $\gtrsim$10$^{32}$~erg~s$^{-1}$ and $\gtrsim$7.5$\times$10$^{32}$~erg~s$^{-1}$ for the point-like Source~1 and extended Source~2, respectively, would also be compatible with typical luminosities and luminosity ratios of pulsar/PWN associations.

Complementary studies are needed  to confirm or reject any of the scenarios  we explored here regarding the point-like and extended diffuse X-ray sources. High-resolution X-ray observations from Chandra satellite and radio observations performed with Parkes observatory are needed to compare X-ray spectra and morphology with those at the radio-continuum bands and to try to detect or put constraints on pulsed emission coming from the point-like source.

\begin{acknowledgements}
FG is a fellow of CONICET. JAC, MCM and GER are CONICET researchers. JAC was supported in different aspects of this work by Consejer\'{\i}a de Econom\'{\i}a, Innovaci\'on, Ciencia y Empleo of Junta de Andaluc\'{\i}a under excellence grant FQM-1343 and research group FQM-322, as well as FEDER funds. GER is supported by grants from ANPCyT (PICT 2012-00878) and Spanish MICINN (AYA 2013-47447-C3-1-P).
\end{acknowledgements}


\end{document}